\documentclass{article}

\newcommand{\ttitle}{Heuristic-Based Weak Learning for Moral Decision-Making}

\newif\ificml
\newif\ifanon
\newif\ifcomments
\icmltrue
\commentstrue
\usepackage{lipsum}
\usepackage{paralist}

\ificml
\else
\usepackage{minted}
\usepackage{hyperref}
\usepackage{natbib}
\usepackage[colorinlistoftodos]{todonotes}
\usepackage{amsmath}
\usepackage{amsfonts}
\DeclareMathOperator*{\argmin}{argmin}

\usepackage{makecell}
\usepackage{dsfont}
\fi
\renewcommand{\ttitle}{Heuristic-Based Weak Learning for Automated Decision-Making}
\usepackage[hyphens]{url}

\usepackage{microtype}
\usepackage{graphicx}
\usepackage{subfigure}
\usepackage{booktabs} 

\usepackage{hyperref}

\usepackage[accepted]{icml/icml2020}
\anonfalse

\icmltitlerunning{\ttitle}

\begin{document}

\twocolumn[
\icmltitle{\ttitle}




\begin{icmlauthorlist}
\icmlauthor{Ryan Steed}{cmu,gwu}
\icmlauthor{Benjamin Williams}{gwu}
\end{icmlauthorlist}

\icmlaffiliation{cmu}{Heinz College, Carnegie Mellon University, Pittsburgh, PA, USA}
\icmlaffiliation{gwu}{Columbian College of Arts and Sciences, George Washington University, Washington, DC, USA}

\icmlcorrespondingauthor{Ryan Steed}{ryansteed@cmu.edu}

\icmlkeywords{Weak learning, moral AI}

\vskip 0.3in
]



\printAffiliationsAndNotice{}  

\begin{abstract}
\ificml
Machine learning systems impact many stakeholders and groups of users, often disparately. Prior studies have reconciled conflicting user preferences by aggregating a high volume of manually labeled pairwise comparisons, but this technique may be costly or impractical. How can we lower the barrier to participation in algorithm design? Instead of creating a simplified labeling task for a crowd, we suggest collecting ranked decision-making heuristics from a focused sample of affected users. With empirical data from two use cases, we show that our weak learning approach, which requires little to no manual labeling, agrees with participants’ pairwise choices nearly as often as fully supervised approaches.

\fi
\end{abstract}

\section{Introduction}
\label{sec:intro}

\ificml

With the widespread use of machine learning (ML) systems comes a growing set of applications where a few stakeholders have disproportionate influence over an algorithm's decisions. Too often, this power imbalance perpetuates harms to specific groups of users, even in government-run decision-making systems \citep{Angwin2016MachineBias, Chouldechova2017FairInstruments, Noble2018AlgorithmsRacism}. In many cases, it is highly desirable for affected users to have input in algorithm design. For problems of this sort, multiple social values must be considered and no one objective can satisfy all stakeholders.

To facilitate stakeholder participation in algorithm design, several studies have attempted to solicit and aggregate users' preferences. These approaches often rely on hypothetical ``trolley" dilemmas: for instance, should a self-driving car with brake failure swerve into a barrier, killing its passengers, to avoid killing a jaywalker \citep{Awad2018TheExperiment, Noothigattu2018AMaking, Kim2018AMaking}? In the kidney exchange market, should a clearing-house algorithm allocate kidneys to patients who drink less, all else equal \citep{Freedman2018AdaptingValues}? Other studies address more concrete applications, such as on-demand food distribution \citep{Kahng2019StatisticalDemocracy, Lee2019WeBuildAI:Governance}.

Notably, all of these studies rely on pairwise comparisons to estimate people's decision criteria, whether through viral online crowd-sourcing with many respondents or extended interaction with a few participants \citep{Alvarado2018TowardsContexts, Zhu2018Value-SensitiveLessons}. There are potentially prohibitive limitations associated with pairwise crowd-sourcing: the economic cost of hand-labeling enough comparisons to train an accurate model is high, especially for scenarios that require rare expertise, and stakeholders may not trust ML models built without discernible rules \cite{Lee2019WeBuildAI:Governance}. On the other hand, ``top-down" rules-based models can be unrealistic and inflexible. Is there a cheaper mechanism for eliciting and aggregating complex stakeholder preferences?

We propose a weak learning approach to stakeholder participation that requires little to no manual labeling. We suggest asking users to report and rank simple rules, or heuristics, that govern their preferences for an system's design. By combining multiple heuristics and considering their popularity among users, we automatically label training data for standard ML models. As a proof-of-concept, we compare our approach to fully supervised learning for two use cases with existing crowd-sourced data. We show that weak learning models agree with respondents' pairwise preferences almost as often as fully supervised models. If there are common heuristics, then ranking the heuristics by popularity results in nearly identical rates of agreement.

\fi

\ificml

\section{Related Work}
\label{sec:background}

Most approaches to automated preference aggregation for decision-making take one of two stances: a ``top-down" design paradigm, in which a set of pre-determined rules govern decision-making, often in the form of short or extensive games \citep{Dehghani2008MoralDM:Decision-Making, Anderson2014GenEth:Analyzer, Blass2015MoralExemplars, Conitzer2017MoralIntelligence}; or a `bottom-up" attitude, where preferences are learned from user data with ML or other statistical methods \citep{Kim2018AMaking, Noothigattu2018AMaking, Kahng2019StatisticalDemocracy, Freedman2018AdaptingValues}. \citet{Lee2019WeBuildAI:Governance} combine elements of both paradigms, asking participants to construct two decision models: an ML model trained on their pairwise preferences, and an explicit rule model in which participants vary the parameters of a fixed objective. Top-down approaches are more transparent to stakeholders, but shift influence towards the system designers who collect and interpret qualitative stakeholder interests. Bottom-up approaches better represent expert preferences in new scenarios, but are susceptible to embedded biases in data and incur additional data collection costs.

\fi

\section{Approach}
\label{sec:approach}

\ificml

We seek to design a mechanism for preference elicitation that is more flexible than abstracted approaches like the explicit rule model, but still accords with stakeholders' concrete, pairwise preferences. Our solution eliminates the manual labeling required in participatory frameworks like the ones above. Rather than collecting a single explicit rule model from each stakeholder representative, or domain expert, we draw on the emerging field of weak learning to aggregate multiple weak heuristics from stakeholders into automatically labeled training data for standard models.

\subsection{Heuristics}
Heuristic functions are practical, usually simplistic, rules for determining the ``right" decision in a given scenario. As an example in the AV trolley problem, a German ethics commission recommended that AVs should place the protection of human life above other animal life, a utilitarian heuristic \citep{Luetge2017TheDriving}. Because of the context-dependent nature of decision-making, individual \emph{weak} heuristics are not expected to be correct all or even most of the time; rather, each heuristic performs fairly well on a subset of the problem. We implement each heuristic in Python using the Snorkel \citep{Snorkel} labeling function interface.


This method of elicitation has already been tested with human participants: \citet{Ratner2017Snorkel:Supervision} trained several domain experts with prior coding experience to write heuristic functions in a two-day workshop.\footnote{It should be noted that some translation, manual or automated, is necessary to survey participants without a coding background.} For a classification problem in the field of bioinformatics, these users achieved better accuracy scores than hand-labelers on Amazon Mechanical Turk (MTurk) by writing only a few heuristics.

\subsection{Workflow}
Our weak learning approach is inspired by \citet{Ratner2017Snorkel:Supervision}, whose open-source Snorkel library aggregates labeling functions to programmatically generate training data for ML models. We propose a general workflow as follows:

\begin{enumerate}
    \item \emph{Heuristic Design:} Develop a fixed number of decision-making heuristics with each participant in the form of plain language or pseudo-code. If possible, ask participants to rank their heuristics according to strength of belief or perceived accuracy.
    \item \emph{Heuristic Tuning:} Write a prototype version of each heuristic as a function over a set of alternatives. Collaboratively ensure that each heuristic function adequately expresses the expert's decision logic.
    \item \emph{Candidate Labeling:} Evaluate all heuristics on a set of pairwise comparisons (like those presented to participants in a crowd-sourcing approach), producing a set of candidate labels with each heuristic function. 
    \item \emph{Label Aggregation:} Aggregate the candidate labels by majority vote, where each heuristic function is a voter, or with a generative model like that of \citet{Bach2017LearningData}. If rankings are available and there are duplicate heuristics, weight each according to the empirical popularity of the heuristic amongst stakeholders. This step produces a single aggregate label for each comparison.
    \item \emph{Discriminative Modeling:} Train or estimate a discriminative model (classifier) on the aggregate labels produced in (4). This step generalizes beyond the aggregate heuristic labels, making the model slightly more flexible. Validate with a held-out test set.
\end{enumerate}

\fi

\newcommand{\mmlmacc}{63.0}
\newcommand{\mmmvacc}{67.9}
\newcommand{\mmvaccweighted}{68.5}
\newcommand{\mmclfgoldacc}{69.8}
\newcommand{\mmclffiltered}{0}
\newcommand{\mmclflmacc}{66.6}
\newcommand{\mmclfmvacc}{66.8}
\newcommand{\mmaccsgoldvoter}{70.0}
\newcommand{\nuniquealt}{2.3}
\newcommand{\mmnumtrials}{50}

\newcommand{\keclfgoldacc}{86.0}
\newcommand{\keclfmvacc}{81.1}
\newcommand{\keclfbordaacc}{85.6}
\newcommand{\kefreedmanacc}{85.8}
\newcommand{\kenumtrials}{50}

\newcommand{\experiment}[1]{\ificml{\emph{#1.}}\fi}

\section{Experiments}
\label{sec:experiments}

\ificml

As a proof-of-concept, we evaluate whether our approach can capture users' pairwise preferences without any manual labeling. To compare our approach to fully supervised learning, we source heuristic functions from data published by the authors of two test datasets: the autonomous vehicle (AV) trolley dilemma \citep{Awad2018TheExperiment} and the kidney exchange \citep{Freedman2018AdaptingValues}. We hypothesize that the heuristic-based model has a comparable rate of agreement with respondent decisions to fully supervised models trained directly on these datasets. This study focuses on binary classification and ethical preferences, but our method is also suitable for more complex applications.

For both test cases, we develop example heuristics based on available ground-truth pairwise comparisons and then train a classifier on a subset of the aggregated heuristic labels. We compare the accuracy of these \emph{weakly supervised} models to \emph{fully supervised} classifiers trained directly on ground-truth responses. Candidate labels are aggregated with a simple majority voting model; ties are broken using weights estimated by the generative model described in \citet{Bach2017LearningData}. For the kidney exchange, we construct an additional voting model that weights each heuristic by its popularity among respondents. For the discriminative model, we use random forest classifiers with 100 estimators, Gini split criterion, no maximum depth, and a minimum of two samples per split. In both use cases, the ground-truth labels are balanced.\footnote{All heuristic functions, data, and code used to produce the figures in this paper are available at \ifanon{\url{https://github.com/anonymous/repo}}\else{\url{https://github.com/ryansteed/heuristic-moral-machine}}\fi.}

\fi

\subsection{Autonomous Vehicle Trolley Problem}
\label{sec:moralmachine}

\ificml

\citet{Awad2018TheExperiment} explore the classic trolley problem with a modern twist: autonomous vehicles. Visitors to the Moral Machine website \citep{MoralMachine} intervene to save one of two different sets of characters, each with a unique set of potentially relevant attributes. To simplify the problem, we follow \citet{Kim2018AMaking} in decomposing the full feature vector into an abstracted one: the number of male, female, young, old, infant, pregnant, fat, fit, working, medical, homeless, criminal, human, non-human, passenger, law-abiding, and law-violating characters saved by the respondents' choice. Of the published data, this experiment considers only votes from complete 13-comparison sessions (totaling 1,544,920 decisions from 51,211 unique respondents).

\begin{figure}[!t]
    \centering
    \includegraphics[width=\linewidth]{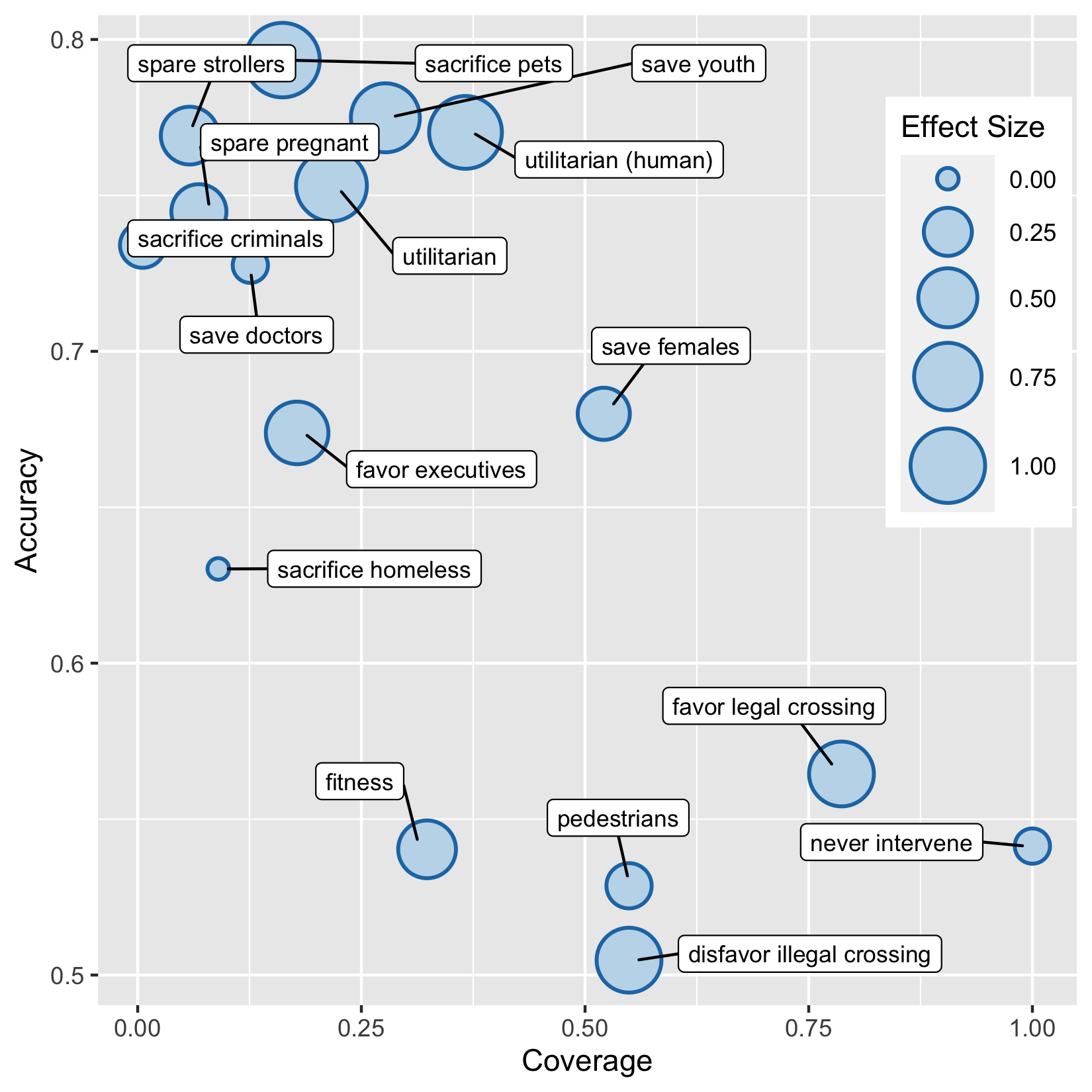}
    \caption{Rate of agreement between heuristic functions and Moral Machine respondents in the validation set. Accuracy is the rate at which which the heuristic agreed with respondents. Coverage is the rate at which the heuristic did not abstain. Heuristics are sized by the corresponding strength of respondent preference, as measured by  \citet{Awad2018TheExperiment}.}
    \label{fig:mm-weights}
\end{figure}

\experiment{Heuristic Accuracy}
As a proof-of-concept, we assume that if asked about their reasoning, respondents would have reported heuristics aligned with the preferences they expressed through pairwise comparisons. Mirroring the cross-cultural preferences estimated by \citet{Awad2018TheExperiment}, we wrote a set of 16 functions expressing heuristics for each of the statistically significant preferences (e.g. ``save doctors" and ``do not hit the pedestrians if they are crossing legally"). Heuristics were debugged using a held-out development partition containing 20\% of responses from the test set. We further partitioned the training set to obtain a validation set with 106,105 responses (20\% of the development set). It takes only seconds to produce candidate labels for each data point. Figure~\ref{fig:mm-weights} shows the accuracy and coverage for each heuristic. Usually, those heuristics which correspond to a strong respondent preference are also the most accurate, with a few exceptions.

\experiment{Discriminative Model Accuracy}
We then trained a discriminative model on the aggregated labels to generalize to new dilemmas. All classifiers in the following section were trained on the training partition (424,419 examples) and tested on a separate test partition of 106,105 dilemmas presented to respondents. We imputed the median for approximately 140,000 dilemmas with missing feature values. Figure \ref{fig:mm-accs_voter} plots the learning curve for both of our models. For comparison, \citet{Kim2018AMaking} measure approximately 75\% out-of-sample prediction accuracy for their hierarchical Bayesian approach to learning moral preferences; \citet{Noothigattu2018AMaking} do not report prediction accuracy for non-synthetic data. Our weak learning model agrees with respondents slightly less (\mmclfmvacc\%) than this statistical model, but agrees with pairwise decisions nearly as often as an identical, fully supervised classifier (\mmclfgoldacc\%).

\begin{figure}[!b]
    \centering
    \includegraphics[width=\linewidth]{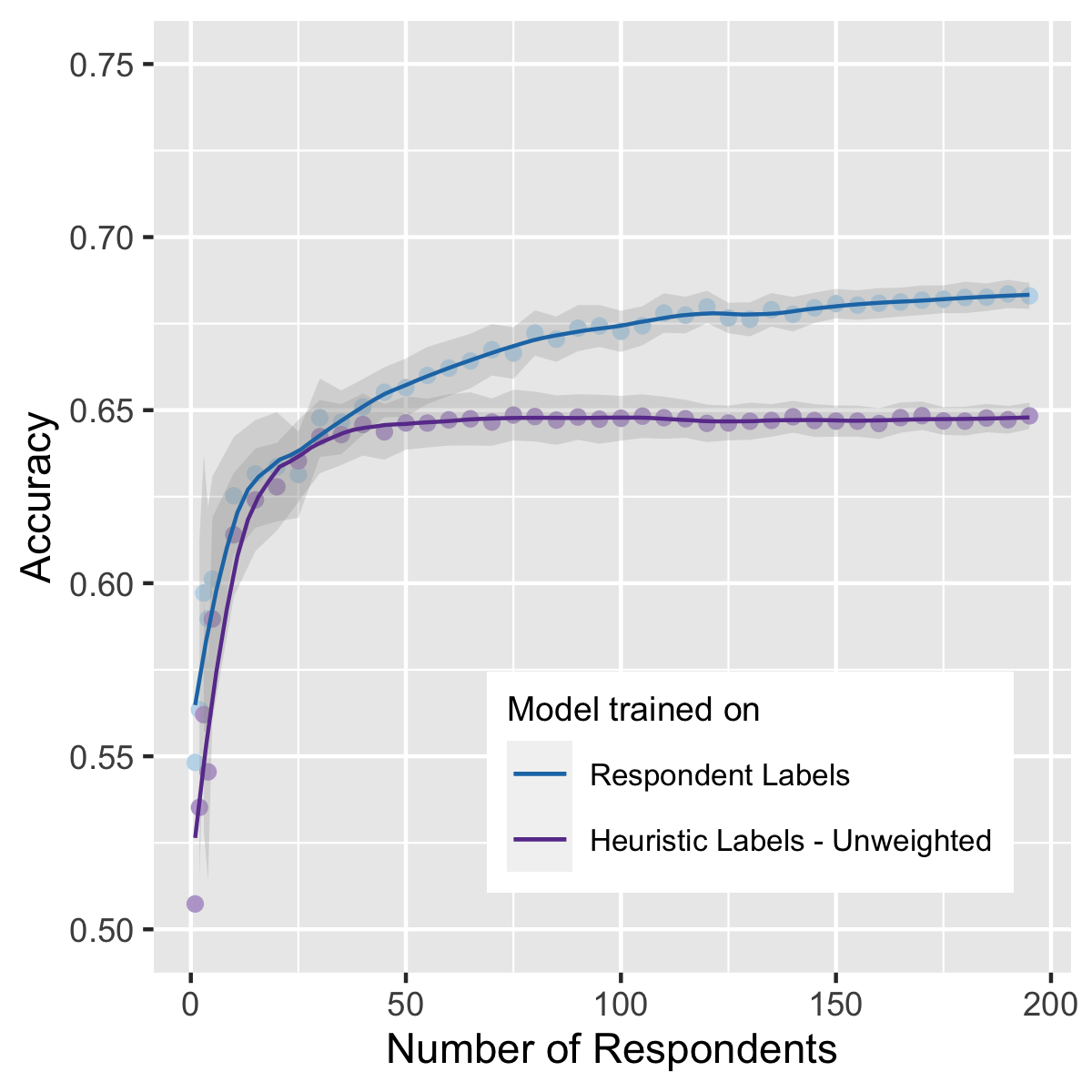}
    \caption{Discriminative model accuracy increase as the number of respondents is increased. The discriminative model and weighted voting model are fitted on a random training partition of $n$ respondents' responses and tested on the rest. 95\% confidence intervals for the mean accuracy across {\kenumtrials} are shown in gray.}
    \label{fig:mm-accs_voter}
\end{figure}

\fi

\subsection{Kidney Exchange}
\label{sec:kidneyexchange}

\ificml

In kidney exchanges, a central market clearing algorithm matches kidney donors to patients in need of an organ. Patients may be prioritized according to a mixture of logistical and moral criteria. \citet{Freedman2018AdaptingValues} asked 289 MTurk users to allocate a kidney in a series of pairwise dilemmas. Each fictional patient is young or old, drinks alcohol rarely or frequently, and may or may not have other health problems. Each respondent was presented with all 28 pairwise contests, for a total of 8,092 comparisons. Each feature was coded as a binary variable.

\experiment{Heuristic Accuracy}
For this study, no global assumption about respondent heuristics is necessary: respondents were asked to explain the strategy they used to decide which patient should receive the kidney. We found three common heuristics: ``choose younger patient" (83.2\% accurate), ``choose patient who drinks less" (78.8\%), and ``choose patient with no other health issues" (61.0\%). Because all permutations of the feature space were included in the survey, heuristic coverage is roughly equal.

\experiment{Heuristic Popularity}
Because this survey included free-form responses, we conducted an additional experiment in which the respondents' reported strategies informed heuristic weighting during aggregation. Respondents usually provided a series of ranked heuristics: e.g., ``I always chose the younger patient; if both patients were the same age, then I chose the one who drank less." We manually coded each qualitative response into a ranking, ties permitted. We omitted 35 respondents who did not provide a strategy or whose strategies did not match any of those listed in Table~\ref{tab:ke-borda}, which reports the average Borda counts for these coded strategies. We then modify the majority voting model to weight the votes from each heuristic function. The voting weights are simply the average Borda counts scaled from 0 to 1.

\begin{table}[!b]
    \caption{Mean Borda count for each heuristic and its contradiction. Borda counts are calculated from manual ranked choice coding of text responses. Ties are permitted.}
    \vskip 0.15in
    \centering\small
    \begin{tabular}{lc}
    \toprule
    Heuristic & Avg. Borda Count  \\
    \midrule
    Choose older patient & 0.11 \\
    Choose younger patient & 3.42 \\
    Choose patient who drinks more & 0.04 \\
    Choose patient who drinks less & 2.71 \\
    Choose patient with other health issues & 0.19 \\
    Choose patient with no other health issues & 2.10\\
    \bottomrule
\end{tabular}
    \label{tab:ke-borda}
\end{table}

\experiment{Discriminative Model Accuracy}
For comparison, \citet{Freedman2018AdaptingValues} designed a kidney exchange algorithm that chooses the patient whose profile commands the higher normalized moral preference, all other factors being equal. Despite sampling all possible permutations of the problem space, their strategy agrees with the MTurk respondents in 10-fold cross validation only about as often as a supervised baseline classifier: {\kefreedmanacc}\% and {\keclfgoldacc}\%, respectively. Again, our weakly supervised approach performs slightly worse, agreeing with respondents {\keclfmvacc}\% of the time. In contrast, the \emph{weighted} majority voting model, which accounts for heuristic popularity, agrees just as often as the fully supervised baseline without any manually labeled data ({\keclfbordaacc}\%). Figure \ref{fig:ke-accs_voter} shows the learning curve for each model. The model trained on popular votes learns at a much quicker rate per number of respondents than even the fully-supervised classifier, and achieves the same equilibrium accuracy. Rather than hand-labeling, participants could produce and vote on a set of heuristics.

\begin{figure}[!t]
    \centering
    \includegraphics[width=\linewidth]{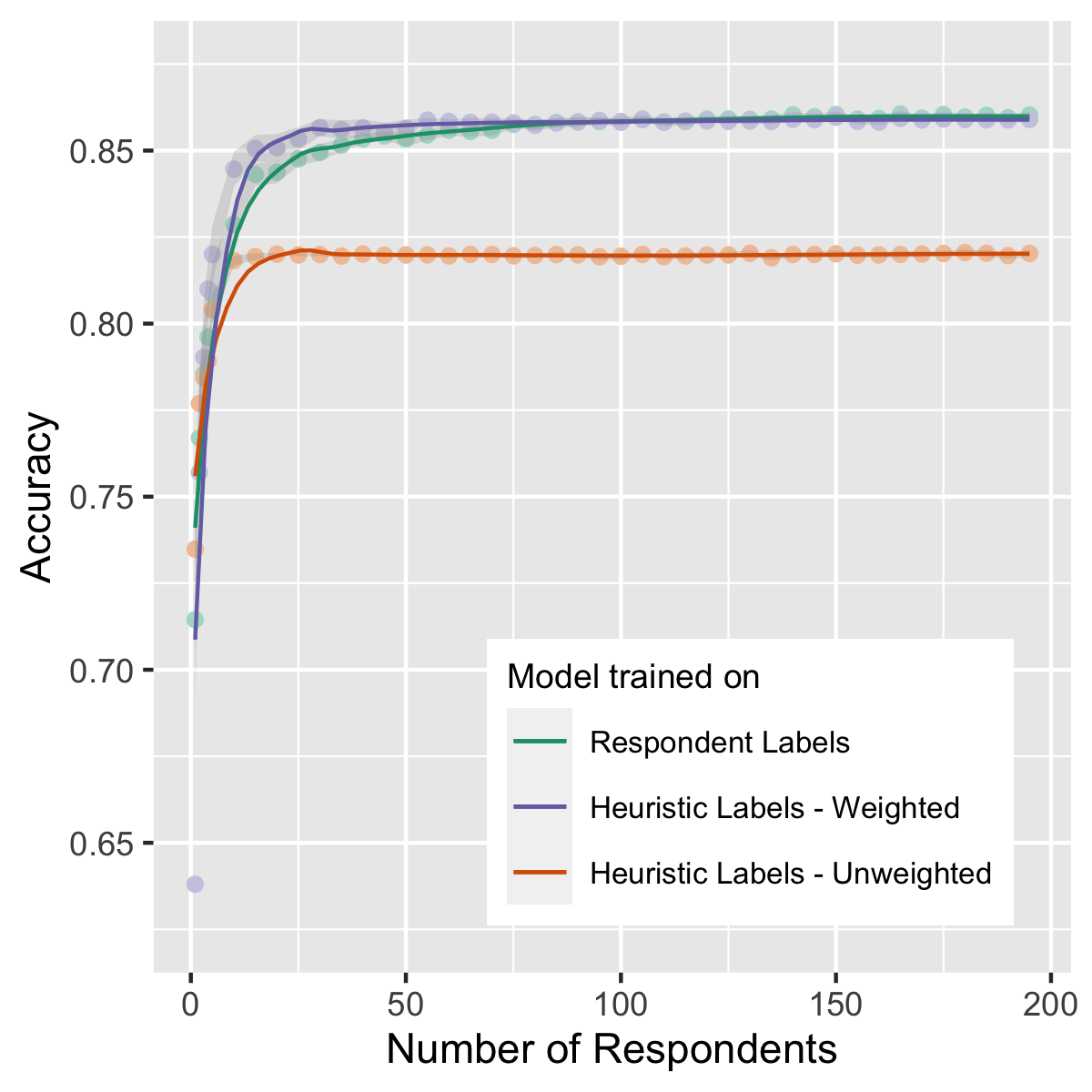}
        \caption{Discriminative model accuracy increase as the number of respondents is increased. The discriminative model and weighted voting model are fitted on a random training partition of $n$ respondents' responses and tested on the rest. 95\% confidence intervals for the mean accuracy across {\kenumtrials} are shown in gray.}
    \label{fig:ke-accs_voter}
\end{figure}

\fi

\section{Conclusion}
\label{sec:conclusion}

\ificml

By lowering the amount of manual labeling required to represent stakeholder interests, this approach lowers the barrier to stakeholder participation in ML systems. Moreover, weak supervision provides the opportunity for more complex expressions of stakeholder preferences. In this paper, we validate our approach against pairwise comparisons, showing that in domains where ground-truth data already exists, heuristic-based learning performs comparably well to crowd-sourcing approaches. Future work will test the advantages of this approach in situations where participants need to express more complicated, overlapping preferences and examine stakeholder trust in heuristic-based models.

\fi

\ifanon
\else
\section{Acknowledgements}
We thank Rahul Simha, Brian Wright, and Rachel Riedner for their insightful comments on an earlier version of this paper. Also, we thank Totte Harinen for pointing out a bug in an earlier version of our code. 
\fi

\bibliography{main.bib}
\bibliographystyle{icml/icml2020.bst}

\end{document}